\documentclass[a4paper]{jpconf}
\usepackage{graphicx}

\newcommand{\be}{\begin{equation}}
\newcommand{\ee}{\end{equation}}
\newcommand{\bea}{\begin{eqnarray}}
\newcommand{\eea}{\end{eqnarray}}

\def\<{\langle}
\def\>{\rangle}
\def\psl{/ \! \! \! p}

\begin{document}
\title{Exact Ward-Takahashi identity for the lattice $N=1$ Wess-Zumino model}

\author{Alessandra Feo\footnote{Report on work with Marisa Bonini.}}

\address{Dipartimento di Fisica, Universit\'a di Parma,
         and INFN Gruppo Collegato di Parma,
         Parco Area delle Scienze 7/A. 43100 Parma, Italy}

\ead{feo@fis.unipr.it}

\begin{abstract}
The lattice Wess-Zumino model written in terms of the Ginsparg-Wilson relation 
is invariant under a generalized supersymmetry transformation which is determined by an iterative procedure in 
the coupling constant. By studying the associated Ward-Takahashi 
identity up to order $g^2$ we show that this lattice supersymmetry automatically leads to restoration of 
continuum supersymmetry without fine tuning. 
In particular, the scalar and fermion renormalization wave functions coincide. 
\end{abstract}

\section{Introduction}
The study of $N=1$ Super Yang-Mills theory on the lattice has been implemented using Wilson fermions 
\cite{montvay} starting from a non-exact lattice supersymmetry. Thus, to recover the continuum supersymmetric
theory a fine tuning is needed (see \cite{feo} for review). To avoid this problem an  
exact formulation of supersymmetry on the lattice would be required. It would protect the 
theory from dangerous SUSY-violating radiative corrections terms and no fine tuning
(see Refs. \cite{various} for different approaches on exact formulations of extended supersymmetries). 

In this report we {\em explicitly} show how supersymmetry is recovered in the continuum limit without fine tuning
when starting from an exact supersymmetry of the lattice action. We prove this result \cite{bf,bf2} in the special
case of the 4-dimensional lattice $N=1$ Wess-Zumino model introduced in Ref. \cite{fujikawa}.

To start, let us write down the lattice $N=1$ Wess-Zumino model in terms of real components as
$S_{WZ}=S_0 + S_{int}$ where
\bea
&& \hspace{-0.7cm} S_0 = \sum_x \bigg \{ \frac{1}{2} \bar \chi (1 - \frac{a}{2} D_1)^{-1} D_2 \chi - 
\frac{1}{a} ( A D_1 A + B D_1 B)  
+ \frac{1}{2} F (1 - \frac{a}{2} D_1)^{-1} F + \frac{1}{2} G (1 - \frac{a}{2} D_1)^{-1} G 
\bigg\} \nonumber \\
&& \hspace{-0.7cm} S_{int} = \sum_x \bigg\{ \frac{1}{2} m \bar \chi \chi + m (F A + G B) 
+ \frac{1}{\sqrt{2}} g \bar \chi (A + i \gamma_5 B) \chi
+ \frac{1}{\sqrt{2}} g \big[ F (A^2 - B^2) + 2 G (A B) \big] \bigg\} 
\eea
and $A,B,F,G$ are the scalar and auxiliary fields while $\chi$ is a Majorana fermion that satisfies the 
Majorana condition 
$\bar \chi = \chi^T C$. $C$ is the charge conjugation matrix that satisfies 
$C^T = -C$ and $C C^\dagger = 1$. Moreover, 
\be
D_1 = \frac{1}{a} \bigg[ 1 - (1 + \frac{a^2}{2} \nabla^\star_\mu \nabla_\mu) \frac{1}{\sqrt{X^\dagger X}} 
\bigg] \, , \qquad \qquad 
D_2 = \frac{1}{2} \gamma_\mu (\nabla^{\star}_\mu + \nabla_\mu) \frac{1}{\sqrt{X^\dagger X}} \equiv  
\gamma_\mu D_{2 \mu} \, ,
\ee
where $X = 1 - a D_w $ \cite{neuberger}. 
In terms of $D_1$ and $D_2$ the Ginsparg-Wilson relation \cite{ginsparg}, 
$\gamma_5 D + D \gamma_5 = a D \gamma_5 D$ (that may be regarded as a lattice form of the chiral 
symmetry \cite{luscher} and protects the fermion masses from additive 
renormalization) becomes,
$D_1^2 - D_2^2 = \frac{2}{a} D_1$ and $(1-\frac a2 D_1)^{-1}D_2^2=-\frac 2a D_1$. 
In the continuum limit $S_{WZ}$ reduces to the continuum $N=1$ Wess-Zumino action. 

In Ref. \cite{bf} we showed that $S_{WZ}$ is invariant under a generalized lattice supersymmetry 
transformation
\bea
&& \delta A = \bar \varepsilon \chi = \bar \chi \varepsilon  \, , 
\qquad \delta B = -i \bar \varepsilon \gamma_5 \chi = -i \bar \chi \gamma_5 \varepsilon \, , 
\qquad  \delta F = \bar \varepsilon D_2 \chi \, , \qquad \delta G = i \bar \varepsilon D_2 \gamma_5 \chi \, ,
\nonumber \\ 
&& \delta \chi = - D_2 (A - i \gamma_5 B) \varepsilon - (F - i \gamma_5 G) \varepsilon + 
g R \varepsilon  \, ,
\label{susy}
\eea
which contains a function $R$ to be determined by imposing $\delta S_{WZ}=0$ 
order by order in $g$. Expanding $R$ in powers of $g$, 
$R = R^{(1)} + g R^{(2)} + \cdots$, we find 
$R^{(1)} = ((1 - \frac{a}{2} D_1)^{-1} D_2 + m )^{-1} \Delta L $ with 
\be
\Delta L \equiv \frac{1}{\sqrt{2}} \Big\{2 (A D_2 A - B D_2 B) - D_2 (A^2 - B^2) 
+ 2 i \gamma_5 \Big[(A D_2 B + B D_2 A) - D_2 (A B)\Big] \Big\} 
\ee
which explicitly shows the breaking of the Leibniz rule at finite lattice spacing.
The function $R$ can be summed up: its formal solution to all orders in $g$ is 
$\big[(1 - \frac{a}{2} D_1)^{-1} D_2  + m  + \sqrt{2} g (A + i \gamma_5 B) \big] R  = \Delta L $. 
Notice that $R \to 0$ when $a \to 0$ since $\Delta L$ vanishes in this limit. 

This generalized supersymmetry transformation (\ref{susy}) satisfy a distorted algebra whose general expression 
for the commutator is given by
$[\delta_1 , \delta_2] \Phi = \alpha^\mu P^\Phi_\mu(\Phi)$ where $\Phi = (A,B,F,G,\chi)$
and $\alpha^\mu = -2 \bar \varepsilon_2 \gamma^\mu \varepsilon_2$. $P^\Phi_\mu(\Phi)$ are polynomials in 
$\Phi$ defined as $P^\Phi_\mu(\Phi) = D_{2 \mu}\Phi + O(g)$. 
We have verified (up to order $g^1$) that the closure works, i.e. the action is invariant under the 
transformation $\Phi \to \Phi + \alpha^\mu P^\Phi_\mu(\Phi)$.
Notice that in the continuum limit $D_{2 \mu} \to \partial_\mu $  and the transformation  
reduces to $\Phi \to \Phi + \alpha^\mu \partial_\mu \Phi $. 

\section{Two-point Ward-Takahashi identity and the continuum limit}
\begin{figure}[ht]
\begin{center}
\includegraphics[width=18pc]{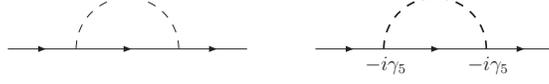}
\end{center}
\caption{\label{xx} Feynman diagrams for the non-tadpole contributions to $\<\chi\bar\chi\>^{(2)}$.}
\end{figure}

Let us study the consequences of this exact generalized lattice supersymmetry. In order to
do so, let us concentrate on some Ward-Takahashi identity (WTi). The WTi is derived from the 
generating functional 
$Z[\Phi,J]=\int{\cal D}\Phi \exp{-(S_{WZ}+S_J)}$
where $S_J$ is the source term $S_J=\sum_x J_\Phi \cdot\Phi
\equiv \sum_x\Bigg\{J_A\,A+J_B\,B+J_F\,F+J_G\,G+\bar\eta\chi\Bigg\}$. 
Using the invariance of both, the Wess-Zumino action and the measure with respect
to the lattice supersymmetry transformation, the WTi reads $\< J_\Phi \cdot \delta \Phi \>_J = 0$.
An interesting and non-trivial WTi is the one that relates the fermion and scalar two-point functions.  
Taking the derivative of $\< J_\Phi \cdot \delta \Phi \>_J = 0 $ 
with respect to $\bar \eta$ and $J_A$ and setting to zero all the sources we have
\be
\<\chi_y\bar\chi_x\>-\<D_{2yz}(A_z - i \gamma_5 B_z)A_x\> - \<(F_y - i \gamma_5  G_y)A_x \> + g \< R_y A_x \> 
= 0 \, .
\ee
This identity is trivially satisfied at tree level using the corresponding propagators:
$ \< A A \> = \< B B\> = -{\cal M}^{-1} (1 - \frac{a}{2} D_1)^{-1}$, 
$\< F F\> = \< G G \> = \frac{2}{a}{\cal M}^{-1}  D_1  =-{\cal M}^{-1} (1 - \frac{a}{2} D_1)^{-1} D_2^2$,  
$\< A F \> = \< B G \> = m \,{\cal M}^{-1}$ and 
$\< \chi \bar \chi \> = ((1 - \frac{a}{2} D_1)^{-1} D_2  + m)^{-1}=
-{\cal M}^{-1} ((1 - \frac{a}{2} D_1)^{-1} D_2 - m)$, 
where ${\cal M} = \Big[ \frac{2}{a} D_1 (1 - \frac{a}{2} D_1)^{-1} + m^2 \Big] $. 
The next non-trivial order is $g^2$ which corresponds to the one-loop corrections and can be written as \cite{bf2}
\be
\<\chi_y\bar\chi_x\>^{(2)}
-\<D_{2yz}(A_z - i \gamma_5 B_z)A_x\>^{(2)} - \< (F_y - i \gamma_5  G_y)A_x \>^{(2)} + g \< R^{(1)}_y A_x \>^{(1)} 
+g^2 \< R^{(2)}_y A_x \>^{(0)} = 0 \,.
\label{wt}
\ee
Applying the Wick expansion to the first term we obtain
\bea
&&\<\chi_y\bar\chi_x\>^{(2)}
=\frac{g^2}{4}
\<\chi_y\bar\chi_x\sum_{zu}
\Big[\bar \chi (A + i \gamma_5 B) \chi+ F (A^2 - B^2) + 2 G A B \Big]_z\nonumber\\
&& \phantom{<\chi_y\bar\chi_x>^{(2)}=\frac{g^2}{4}<\chi_y\bar\chi_x}
\times
\Big[\bar \chi (A + i \gamma_5 B) \chi+ F (A^2 - B^2) + 2 G A B \Big]_u
\>^{(0)}\,.
\eea
Using the relations $\mbox{Tr}\<\chi\gamma_5\bar\chi\>=0$ and 
$\mbox{Tr}\<\chi\bar\chi\>=4\<AF\>=4\<GB\>$  we showed that the tadpole 
contributions cancel out.
This property is general and will hold for the other terms in the WTi.
Therefore, one is left with the connected non tadpoles diagrams (see fig.\ref{xx})  
\be
\<\chi_y\bar\chi_x\>^{(2)}_{NT}
=2g^2\sum_{uz}\Big\{
\<\chi_y\bar\chi_z\>\<\chi_z\bar\chi_u\>\<\chi_u\bar\chi_x\>\<A_zA_u\>-
\<\chi_y\bar\chi_z\>\gamma_5\<\chi_z\bar\chi_u\>\gamma_5\<\chi_u\bar\chi_x\>\<B_zB_u\>
\Big\} \, .
\ee

\begin{figure}[ht]
\begin{center}
\includegraphics[width=18pc]{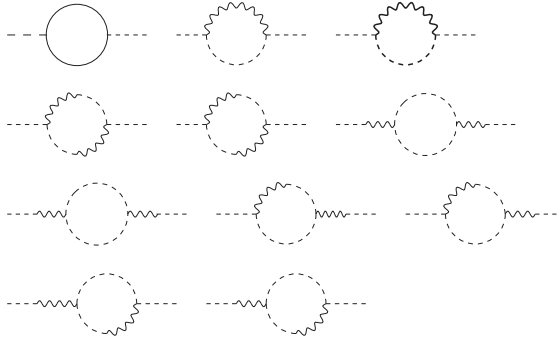}
\end{center}
\caption{\label{aa2}Non-tadpole contributions to $\<D_{2}(A-i\gamma_5 B)A\>^{(2)}$.}
\end{figure}

The non-tadpole contributions to the second term of the WTi are (see fig.\ref{aa2}) 
\bea
&&\<D_{2yz}(A_z - i \gamma_5 B_z)A_x\>^{(2)}_{NT}=g^2\Big\{
D_{2yz}\<A_zA_u\>\Big[\mbox{Tr}\Big(\<\chi_u\bar\chi_w\>\<\chi_w\bar\chi_u\> \Big)
+2\<A_uA_w\>\<F_uF_w\>
\nonumber\\&&
\phantom{\<D_{2yz}(A_z - i \gamma_5}
+2\<B_uB_w\>\<G_uG_w\>
+2\<F_uA_w\>\<A_uF_w\>+2 \<B_uG_w\>\<G_uB_w\>
\Big] \<A_wA_x\>
\nonumber\\&&
\phantom{\<D_{2yz}(A_z - i \gamma_5}
+D_{2yz}\<A_zF_u\>\Big[\<A_uA_w\>\<A_uA_w\>+\<B_uB_w\>\<B_uB_w\>\Big] \<F_wA_x\>
\nonumber\\&&
\phantom{\<D_{2yz}(A_z - i \gamma_5}
+2D_{2yz}\<A_zF_u\>\Big[\<A_uA_w\>\<A_uF_w\>-\<B_uB_w\>\<B_uG_w\>\Big] \<A_wA_x\>
\nonumber\\&&
\phantom{\<D_{2yz}(A_z - i \gamma_5}
+2D_{2yz}\<A_zA_u\>\Big[\<A_uA_w\>\<F_uA_w\>-\<B_uB_w\>\<G_uB_w\>\Big] \<F_wA_x\>\Big\} \, .
\eea

\begin{figure}[ht]
\begin{center}
\includegraphics[width=18pc]{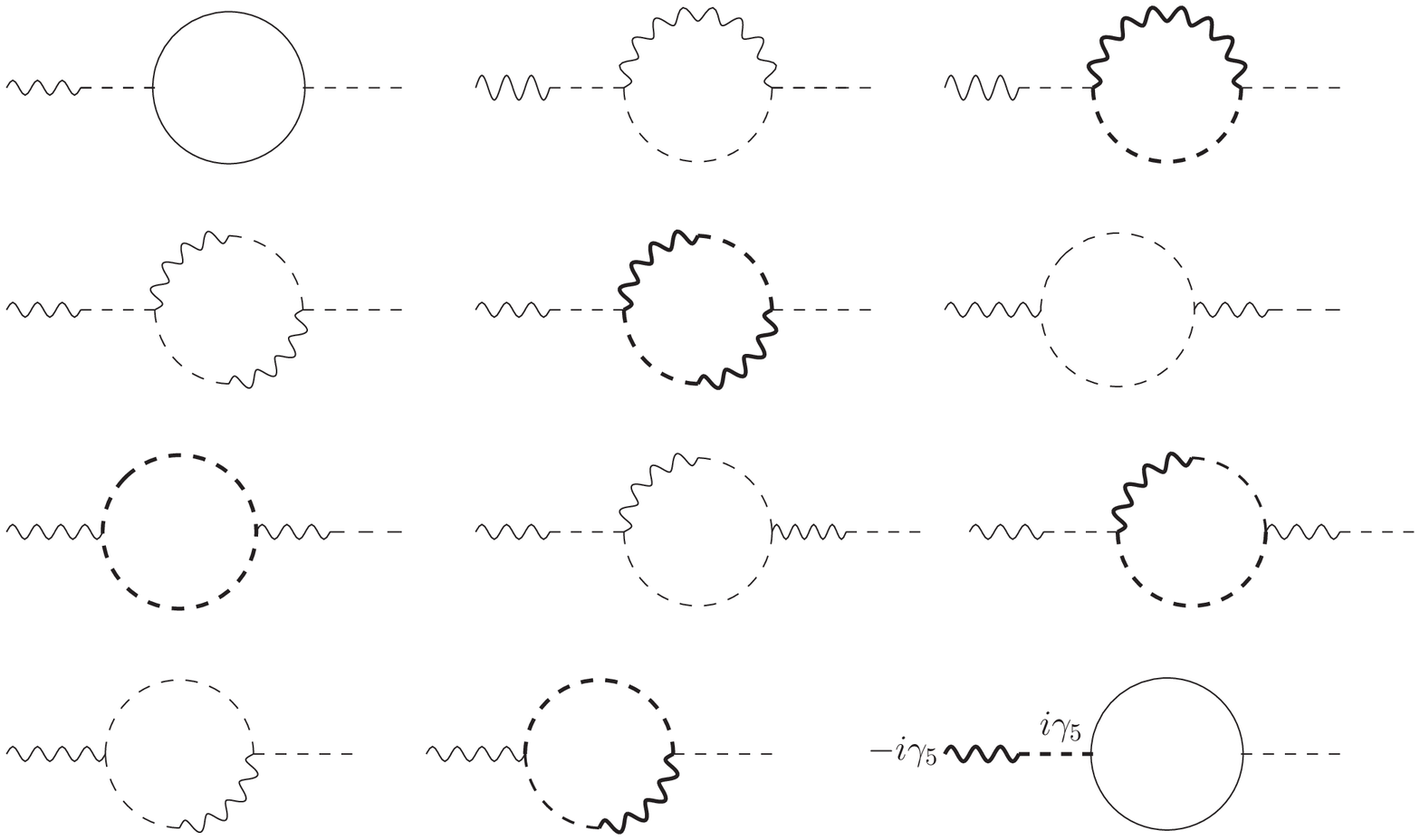}
\end{center}
\caption{\label{fa2}Non-tadpole contributions to $\< (F - i \gamma_5 G) A \>^{(2)}$.}
\end{figure}

The non-tadpole contributions to the third term of WTi are (see fig.\ref{fa2}) 
\bea
&& \< (F_y - i \gamma_5 G_y) A_x \>^{(2)}_{NT}=  g^2\Big\{2\< F_y A_u \> \Big[ 
\frac12 \mbox{Tr} \Big(\<\chi_u\bar\chi_w\> \<\chi_w\bar\chi_u\> \Big) + 
\< F_u F_w \> \< A_u A_w \> \nonumber \\
&& \phantom{ \< (F_y - i \gamma_5 G_y) A_x \>^{(2)}_{NT} }
+ \< F_u A_w \> \< A_u F_w \> + \< G_u G_w \> \< B_u B_w \> + 
\< B_u G_w \> \< G_u B_w \> \Big] \< A_w A_x\> \nonumber \\ 
&& \phantom{ \< (F_y - i \gamma_5 G_y) A_x \>^{(2)}_{NT} }
+ \< F_y F_u \> \Big[ \< A_u A_w \> \< A_u A_w \> 
+ \< B_u B_w \> \< B_u B_w \> \Big] \< F_w A_x \>  \nonumber \\
&& \phantom{ \< (F_y - i \gamma_5 G_y) A_x \>^{(2)}_{NT} }
+ 2 \< F_y A_u \> \Big[ \< F_u A_w \> \< A_u A_w \> 
- \< G_u B_w \> \< B_u B_w \> \Big] \< F_w A_x \>  
\nonumber \\
&& \phantom{ \< (F_y - i \gamma_5 G_y) A_x \>^{(2)}_{NT} }
+ 2 \< F_y F_u \> \Big[ \< A_u A_w \> \< A_u F_w \> 
- \< B_u B_w \> \< B_u G_w \> \Big] \< A_w A_x\> \nonumber \\
&& \phantom{ \< (F_y - i \gamma_5 G_y) A_x \>^{(2)}_{NT} }
- \gamma_5 \< G_y B_w \> \mbox{Tr} \Big( \gamma_5 \< \bar \chi_w \chi_u \> 
\< \bar \chi_u \chi_w \> \Big)  \<A_u A_x \>\Big\} \, . 
\eea

\begin{figure}[ht]
\begin{center}
\includegraphics[width=18pc]{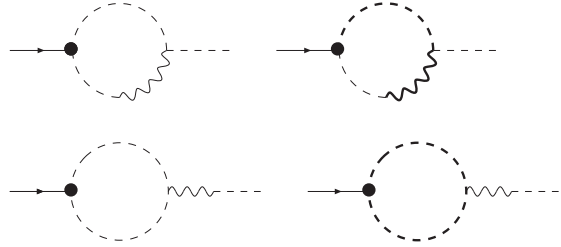}
\end{center}
\caption{\label{r1bis}Non-tadpole contributions to $\< R^{(1)} A\>^{(1)}$. The blob 
denotes the insertion of the operator $D_2$.}
\end{figure}

For the terms of the WTi involving $R$ we find (see fig.\ref{r1bis})
\bea
&& \hspace{-0.8cm}
\< R_y^{(1)} A_x \>^{(1)}_{NT} = -g\<\chi\bar\chi\>_{yz} 
\times\Big\{2 
 \Big[  \<A_z F_w\> D_{2zu} \< A_u A_w\> + \<A_z A_w\> D_{2zu} \< A_u F_w\> 
-D_{2zu}  \<A_u F_w\> \< A_u A_w\> 
\nonumber \\
&&\hspace{-0.7cm} \phantom{ \< R_y^{(1)} A_x\>^{(1)}_{NT} } 
- \<B_z G_w\> D_{2zu} \< B_u B_w\> -  \<B_z B_w\> D_{2zu} \< B_u G_w\> 
+D_{2zu}  \<B_u B_w\> \< B_u G_w\> 
\Big] \< A_w A_x \> 
\nonumber \\ 
&& \hspace{-0.7cm} \phantom{ \< R_y^{(1)} A_x\>^{(1)}_{NT} } 
+ \Big[ 2 \<A_z A_w\> D_{2zu} \< A_u A_w\> 
-D_{2zu}  \<A_u A_w\> \< A_u A_w\> 
\nonumber \\ && 
\hspace{-0.7cm} \phantom{ \< R_y^{(1)} A_x\>^{(1)}_{NT}+\Big[ } 
+ 2\<B_z B_w\> D_{2zu} \< B_u B_w\> 
-D_{2zu} \<B_u B_w\> \< B_u B_w\> 
\Big] \< F_w A_x \> 
\Big\}\,.
\eea

\begin{figure}[ht]
\begin{center}
\includegraphics[width=18pc]{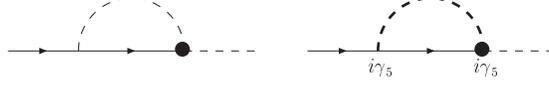}
\end{center}
\caption{\label{r2}Non-tadpole contributions to $\< R^{(2)} A\>^{(0)}$.}
\end{figure}

The last term of WTi is (see fig.\ref{r2}) 
\bea
&&\< R_y^{(2)} A_x \>^{(0)} =
- \sqrt{2}\<\chi\bar\chi\>_{yz} 
\< (A_z+i\gamma_5B_z) \<\chi\bar\chi\>_{zw} \Delta L_w A_x\>^{(0)}
\nonumber\\ 
&&
\hspace{-1.4cm}
= - 2\Big\{
\<\chi_y\bar\chi_z\> \<\chi_z\bar\chi_w\>\Big[\<A_zA_w\>D_{2wu}\<A_uA_x\>
+\<A_wA_x\>D_{2wu}\<A_zA_u\>
-D_{2wu}\<A_zA_u\>\<A_uA_x\>\Big]
\nonumber\\ &&\hspace{-1.4cm}
-\<\chi_y\bar\chi_z\> \gamma_5\<\chi_z\bar\chi_w\> \gamma_5\Big[\<B_zB_w\>D_{2wu}\<A_uA_x\>
+\<A_wA_x\>D_{2wu}\<B_zB_u\>
-D_{2wu}\<B_zB_u\>\<A_uA_x\>\Big] \Big\} \, .
\eea

In order to verify Eq. (\ref{wt}) it is convenient to work in the momentum space representation; then
we can verify that Eq. (\ref{wt}) is exactly satisfied at fixed lattice spacing \cite{bf2}. 
As a last point, let us study the limit $a \to 0$ of Eq. (\ref{wt}) and discuss how this Eq. looks like 
in the continuum limit, how continuum supersymmetry is restored and the role of the operator $\< R A \>$.
Following the notation of Ref. \cite{KY} the continuum limit of the fermion two point function reads \cite{bf2}
\be
\<\chi \bar\chi \>^{(2)}(p)
= \frac{(i \psl{} - m)}{(p^2 + m^2)} C_2 i \psl{} \frac{(i \psl{}  - m)}{(p^2 + m^2)}
\ee
where 
\be
C_2 =g^2  \int \frac{d^4k}{(2 \pi)^4} \frac{\omega'_k \sin^2(k_\rho)}{[(\omega + b)'_k +\frac{ a^2 m^2}{4} 
(\omega - b)'_{k}]^3} + C_{2 f}  
\ee
and $C_{2 f}$ is a finite number while
$\omega'_k\equiv a\omega(k/a) =  \bigg[  1 - 4 \sum_\mu \sin^4(\frac{k_\mu }{2}) + 
4 \bigg(\sum_\mu \sin^2(\frac{k_\mu }{2})\bigg)^2  \bigg]^{1/2}$ and 
$b'_k\equiv \bigg[ \sum_\mu 2 \sin^2(\frac{k_\mu}{2}) - 1 \bigg]$.
For the scalar two point function we obtain 
\be
D_2 \< A  A \>^{(2)}(p) = 
i \psl{} \frac{1}{(p^2 + m^2)} \frac{1}{(p^2 + m^2)} (\frac12 C_3 m^2 - C_1 p^2)
\ee
where 
\be
C_3 =g^2 \int \frac{d^4k}{(2 \pi)^4} \frac{(\omega'_k)^2}{[(\omega' + b')_k 
+ \frac{a^2 m^2}{4} (\omega' - b')_k]^2 } \,,
\ee

\be
C_1 =g^2\int \frac{d^4k}{(2 \pi)^4} \frac{\sin^2(k_\rho)\cos(k_\rho)}
{[(\omega' + b')_k + \frac{a^2 m^2}{4} (\omega' - b')_k]^3}+C_{1f} 
\ee
and $C_{1f}$ is a finite constant. 
A similar analysis applied as before gives 
\be
\< F A \>^{(2)}(p) = 
m \frac{1}{(p^2 + m^2)} \frac{1}{(p^2 + m^2)} (\frac12 C_3+C_1) p^2 \, .
\ee
The continuum limit of the two point function containing the operator $R$ are 
\be
\< R^{(1)} A \>^{(1)}(p) = m \frac{(i \psl{} - m)}{(p^2 + m^2)} \frac{1}{(p^2 + m^2)}
(C_2 -\frac12 C_3)  i \psl{}  
\ee
and
\be
\< R^{(2)} A\>^{(0)}(p) = \frac{(i \psl{}  - m)}{(p^2 + m^2)} \frac{1}{(p^2 + m^2)} (C_2 - C_1)
p^2 \,.
\ee
The combinations $C_2-C_1$ and $C_2-\frac12 C_3$ 
are two (different) finite numbers.  Indeed, the $\log(a^2m^2)$ contributions cancels out
in these combinations. 

Substituting all terms in Eq. (\ref{wt}) with the corresponding signs we have 
\bea
&& \frac{(i \psl{} - m)}{(p^2 + m^2)} ( i \psl{}  C_2) \frac{(i \psl{}  - m)}{(p^2 + m^2)} 
- \frac{i \psl{}  }{(p^2 + m^2)} ( \frac12 m^2 C_3 - p^2 C_1) \frac{1 }{(p^2 + m^2)}  \nonumber \\
&& - \frac{m}{(p^2 + m^2)} (C_1 + \frac12 C_3) p^2  \frac{1 }{(p^2 + m^2)}
+ \frac{(i \psl{}  - m)}{(p^2 + m^2)} (i \psl{}  m) (C_2 - \frac12 C_3) \frac{1 }{(p^2 + m^2)} \nonumber \\
&& + \frac{(i \psl{}  - m)}{(p^2 + m^2)} (C_2 - C_1) p^2 \frac{1 }{(p^2 + m^2)} =0 \, .
\eea
Notice that the pieces coming from the term $\< R A \>$ above are
essential to satisfy the WTi. Thanks to the exactness of WTi  
it is always possible to write the two point function $\<R A \>$ 
as a suitable combination of the other three two point functions involved 
in this WTi. In particular, in the continuum limit one can write
\be
\<R A\>=
\frac{i\psl{} -m}{p^2+m^2}i\psl{} \delta_1\frac{i\psl{} -m}{p^2+m^2}
+i\psl{}  \frac{1}{p^2+m^2}(\delta_2 p^2+\delta_3 m^2)\frac{1}{p^2+m^2} 
-\frac{m}{p^2+m^2}(\delta_2-\delta_3) p^2 \frac{1}{p^2+m^2}
\ee
where $\delta_1=\frac12C_3 -C_2-\delta_3$ and $\delta_2=\frac12C_3-C_1-\delta_3$,
and the constant $\delta_3$ is arbitrary. Then in the continuum limit one 
can rewrite the WTi as the supersymmetric continuum WTi \cite{bf2}
\be
 \< \chi \bar \chi \>_{R}^{(2)} - i\psl{}  \< A A \>_{R}^{(2)} - \< F A \>_{R}^{(2)} = 0 
\ee
with 
$\< \chi \bar \chi \>_{R}^{(2)} \equiv \< \chi \bar \chi \>^{(2)} +
\frac{i\psl{} -m}{p^2+m^2}i\psl{} \delta_1\frac{i\psl{} -m}{p^2+m^2}$, 
$\< A A \>_{R}^{(2)} \equiv\< A A \>_{(2)}
-\frac{1}{p^2+m^2}(\delta_2 p^2+\delta_3 m^2)\frac{1}{p^2+m^2} $ and 
$\< F A \>_{R}^{(2)}\equiv \< F A \>^{(2)}+\frac{m}{p^2+m^2}
(\delta_2-\delta_3)p^2 \frac{1}{p^2+m^2}$. 
It is convenient to express these two point functions in terms of the 1PI vertex functions
(just because we started from an off-shell formulation)
$\< \chi \bar \chi \>^{(2)}=
\frac{i\psl{} -m}{p^2+m^2}\,\Sigma^{(2)}_{\chi\bar\chi}\,\frac{i\psl{} -m}{p^2+m^2}$,
$\< A A\>^{(2)}=
-\frac{1}{p^2+m^2}(\Sigma^{(2)}_{AA}+m^2\Sigma^{(2)}_{FF})\frac{1}{p^2+m^2}$ and 
$\< F A\>^{(2)}=
\frac{1}{p^2+m^2}(\Sigma^{(2)}_{AA}-p^2\Sigma^{(2)}_{FF})\frac{m}{p^2+m^2}$.

The lattice contribution to these 1PI vertices in the continuum limit reads
$\Sigma^{(2)}_{\chi\bar\chi}=i\psl{}   C_2 $, $\Sigma^{(2)}_{AA}=p^2C_1$ and 
$\Sigma^{(2)}_{FF}=-\frac12C_3$.
Moreover, one has 
$\Sigma^{(2)}_{\chi\bar\chi \,R}\equiv
\Sigma^{(2)}_{\chi\bar\chi}+i\psl{} \delta_1=i\psl{} (\frac{C_3}{2}-\delta_3)
\equiv-  Z_{\chi} i\psl{}  $, 
$\Sigma^{(2)}_{AA\,R}\equiv\Sigma^{(2)}_{AA}+p^2\delta_2=
p^2( \frac{C_3}{2}-\delta_3)\equiv-  Z_{A} p^2$ and 
$\Sigma^{(2)}_{FF\,R}\equiv\Sigma^{(2)}_{FF}+\delta_3=
-( \frac{C_3}{2}-\delta_3)\equiv Z_{F}$, 
with 
\be
Z_{\chi}=Z_{A}=Z_{F}=-(\frac{C_3}{2}-\delta_3) \, .
\ee
In the formulation of Fujikawa (without the $R$), the two-point functions of $A$, $F$ and $\chi$
have the same logarithmic divergent parts \cite{fujikawa2} but they differ from different finite 
contributions.  
An important consequence of the exact lattice supersymmetry 
we have introduced is that automatically leads to restoration of supersymmetry in the continuum 
limit with equal renormalization wave function for the scalar and fermion fields.

\end{document}